\documentclass[sigconf, nonacm, authorversion]{acmart}

\usepackage{balance}

\usepackage{hyperref}%

\graphicspath{{figures/}}

\usepackage{tcolorbox}

\usepackage{float}
\newfloat{lstfloat}{htbp}{lop}
\floatname{lstfloat}{Listing}

\usepackage{listings}
\usepackage{pifont}%
\newcommand{\cmark}{\ding{51}}%
\newcommand{\xmark}{\ding{55}}%

\usepackage[inline]{enumitem} %
\usepackage{enumerate}%

\usepackage{multirow}
\usepackage{multicol}

\lstdefinelanguage[ECMAScript2015]{JavaScript}[]{JavaScript}{
	morekeywords=[1]{await, async, case, catch, class, const, default, do,
			enum, export, extends, finally, from, implements, import, instanceof,
			let, static, super, switch, throw, try},
	morestring=[b]` %
}

\lstdefinelanguage{JavaScript}{
	morekeywords=[1]{break, continue, delete, else, for, function, if, in,
			new, return, this, typeof, var, void, while, with},
	morekeywords=[2]{false, null, true, boolean, number, undefined,
			Array, Boolean, Date, Math, Number, String, Object},
	morekeywords=[3]{eval, parseInt, parseFloat, escape, unescape},
	sensitive,
	morecomment=[s]{/*}{*/},
	morecomment=[l]//,
	morecomment=[s]{/**}{*/}, %
	morestring=[b]',
	morestring=[b]"
}[keywords, comments, strings]

\lstalias[]{ES6}[ECMAScript2015]{JavaScript}

\definecolor{mediumgray}{rgb}{0.3, 0.4, 0.4}
\definecolor{mediumblue}{rgb}{0.0, 0.0, 0.8}
\definecolor{forestgreen}{rgb}{0.13, 0.55, 0.13}
\definecolor{darkviolet}{rgb}{0.58, 0.0, 0.83}
\definecolor{royalblue}{rgb}{0.25, 0.41, 0.88}
\definecolor{crimson}{rgb}{0.86, 0.8, 0.24}

\lstdefinestyle{JSES6Base}{
	backgroundcolor=\color{white},
	basicstyle=\ttfamily,
	breakatwhitespace=false,
	breaklines=false,
	captionpos=b,
	columns=fullflexible,
	commentstyle=\color{black}\upshape,
	emph={},
	emphstyle=\color{crimson},
	extendedchars=true,  %
	fontadjust=true,
	frame=single,
	identifierstyle=\color{black},
	keepspaces=true,
	keywordstyle=\color{mediumblue},
	keywordstyle={[2]\color{darkviolet}},
	keywordstyle={[3]\color{royalblue}},
	numbers=left,
	numbersep=5pt,
	numberstyle=\scriptsize\color{black},
	rulecolor=\color{black},
	showlines=true,
	showspaces=false,
	showstringspaces=false,
	showtabs=false,
	stringstyle=\color{forestgreen},
	tabsize=2,
	title=\lstname,
	upquote=true  %
}

\lstdefinestyle{JavaScript}{
	language=JavaScript,
	style=JSES6Base
}
\lstdefinestyle{ES6}{
	language=ES6,
	style=JSES6Base
}

\begin{document}
\title{Confidential Consortium Framework: Secure Multiparty Applications with Confidentiality, Integrity, and High Availability}

\author{Heidi Howard}
\orcid{0000-0001-5256-7664}
\affiliation{%
	\institution{Azure Research, Microsoft}
	\city{}
	\country{}
}
\authornote{Corresponding author's email: \href{mailto:heidi.howard@microsoft.com}{heidi.howard@microsoft.com}}

\author{Fritz Alder}
\orcid{0000-0001-9607-7798}
\affiliation{%
	\institution{imec-DistriNet, KU Leuven}
	\city{}
	\country{Belgium}
}
\authornote{Work done while at Microsoft}

\author{Edward Ashton}
\orcid{0009-0008-7384-2336}
\affiliation{%
	\institution{Azure Research, Microsoft}
	\city{}
	\country{}
}

\author{Amaury Chamayou}
\affiliation{%
	\institution{Azure Research, Microsoft}
	\city{}
	\country{}
}

\author{Sylvan Clebsch}
\affiliation{%
	\institution{Azure Research, Microsoft}
	\city{}
	\country{}
}

\author{Manuel Costa}
\affiliation{%
	\institution{Azure Research, Microsoft}
	\city{}
	\country{}
}

\author{Antoine Delignat-Lavaud}
\affiliation{%
	\institution{Azure Research, Microsoft}
	\city{}
	\country{}
}

\author{C\'{e}dric Fournet}
\affiliation{%
	\institution{Azure Research, Microsoft}
	\city{}
	\country{}
}

\author{Andrew Jeffery}\authornotemark[2]
\orcid{0000-0003-0440-0493}
\affiliation{%
	\institution{University of Cambridge}
	\city{}
	\country{UK}
}

\author{Matthew Kerner}
\affiliation{%
	\institution{Microsoft}
	\city{}
	\country{}
}

\author{Fotios Kounelis}\authornotemark[2]
\orcid{0009-0005-5552-8974}
\affiliation{%
	\institution{Imperial College London}
	\city{}
	\country{UK}
}

\author{Markus A. Kuppe}
\orcid{0000-0002-6972-2031}
\affiliation{%
	\institution{Microsoft Research}
	\city{}
	\country{}
}

\author{Julien Maffre}
\affiliation{%
	\institution{Azure Research, Microsoft}
	\city{}
	\country{}
}

\author{Mark Russinovich}
\affiliation{%
	\institution{Microsoft}
	\city{}
	\country{}
}

\author{Christoph M. Wintersteiger}
\orcid{0000-0003-0102-4381}
\affiliation{%
	\institution{Azure Research, Microsoft}
	\city{}
	\country{}
}

\begin{abstract}
	Confidentiality, integrity protection, and high availability, abbreviated to CIA, are essential properties for trustworthy data systems. The rise of cloud computing and the growing demand for multiparty applications however means that building modern CIA systems is more challenging than ever. In response, we present the Confidential Consortium Framework (CCF), a general-purpose foundation for developing secure stateful CIA applications. CCF combines centralized compute with decentralized trust, supporting deployment on untrusted cloud infrastructure and transparent governance by mutually untrusted parties.
	
	CCF leverages hardware-based trusted execution environments for remotely verifiable confidentiality and code integrity. This is coupled with state machine replication backed by an auditable immutable ledger for data integrity and high availability. CCF enables each service to bring its own application logic, custom multiparty governance model, and deployment scenario, decoupling the operators of nodes from the consortium that governs them. CCF is open-source and available now at \url{https://github.com/microsoft/CCF}.
\end{abstract}

\maketitle

\section{Introduction \& Motivation}\label{sec:introduction}

Together data confidentiality, integrity protection, and high availability form the CIA triad, a classic formulation in information security~\cite{whitman11}. We begin by examining each property, with a particular focus on running trustworthy multiparty applications on untrusted infrastructure.

\emph{\textbf{Data confidentiality.}}
Organizations are responsible for ensuring the privacy of personal data. Increasingly this responsibility is codified in law,
and the penalties for failing to fulfill this obligation can be substantial, for instance up to 4\% of turnover in the case of GDPR~\cite{GDPR}. Even where data is not personal, companies may wish to keep data confidential to protect intellectual property, for competitive advantage, or to maintain systems security, for instance when storing secrets.
Encryption at rest and in-flight are well-established approaches to achieving confidentiality, but confidentiality during execution is more challenging. Moreover, encryption alone does not fully solve the problem of confidentiality. Instead, it reduces the problem of protecting arbitrary data into the problem of protecting keys, which in turn need to be managed, stored, and released according to some well-defined policy.

\emph{\textbf{Integrity protection.}}
Not only are organizations responsible for maintaining data confidentiality, they are also obligated to protect data in their care from unauthorized or unintended modification. In fact, integrity protection of the code that accesses data is often a prerequisite of data confidentiality. Code integrity together with transparency~\cite{Delignat23} allows parties who share data to agree on how data can be used, for instance, a bank can comply with anti-money laundering legislation by executing queries on behalf of a government without disclosing all customer information.

Cloud computing enables efficient scaling of applications with cost proportional to use and a low barrier to entry~\cite{Armbrust10}, but the mass migration to the cloud means that the trusted computing base (TCB) of our systems is growing over time~\cite{Russinovich21}. Data integrity and confidentiality are vital when leveraging untrusted cloud infrastructure~\cite{Singh21} but even more difficult to ensure remotely. As a result, some highly sensitive applications, such as medical, financial, or governmental services, are unable to migrate to the public cloud. This challenging situation leaves us with the following research question: \emph{Is it possible to remove cloud providers from the TCB of multiparty applications and still enable developers to leverage the cloud's compute and storage capabilities?}
This requirement is particularly important with the increasing demand for multiparty scenarios: the integration of data systems between parties who wish to compute over shared data but do not fully trust one another. Collaborating and combining data from multiple sources increases its value and enables novel use cases~\cite{Russinovich21}. However, this makes confidentiality and integrity more difficult, as we must reason about the access rights and requirements of multiple distinct participants.

\emph{\textbf{High availability.}}
As modern society increasingly depends on digital infrastructure, it is vital that applications are dependable and highly available. Even if the necessary consistency and cost trade-offs are made, it is not possible to guarantee 100\% availability of digital infrastructure, and so applications should be resilient to failures expected during normal operation.

\emph{\textbf{CIA triad.}}
We need a principled approach to developing CIA applications that is also highly pragmatic, supporting a wide range of stateful applications and modern deployment scenarios, in particular, delegation to untrusted cloud infrastructure and governance by multiple untrusted parties.

In this paper, we present the Confidential Consortium Framework, also known as CCF, which combines centralized cloud compute with decentralized trust (\autoref{sec:overview}). CCF leverages cloud-based trusted execution environments for remotely attestable confidentiality and integrity (\autoref{sec:intranode}). This is coupled with state machine replication and a transactional key-value store, backed by an immutable ledger for high availability (\autoref{sec:internode}) and auditing. CCF decouples the operators of CCF nodes and the controllers of the system (the consortium members) (\autoref{sec:governance}). CCF is highly flexible, allowing developers to bring their own application logic, and their own multiparty governance model for highly tunable oversight.

As we examine in \autoref{sec:related}, we are not the first to consider data confidentiality, integrity protection, or high availability in the context of cloud computing or multiparty collaboration. Whereas most prior systems offer either a data storage solution (in the form of a ledger, database, or key-value store) or a secure execution solution (relying instead on a secondary storage system) in isolation, CCF offers an end-to-end solution, supporting both execution and storage. CCF has a unique auditable governance model, designed to operate between untrusted environments and parties, codified in a programmable contract. Moreover, CCF occupies a security versus usability sweet spot, offering a relatively small trusted computing base and a simple yet flexible programming model. Last but not least, CCF is trusted in production with services such as Azure Confidential Ledger~\cite{ACL} and Azure Managed CCF~\cite{mCCF} relying on features such as reconfiguration, disaster recovery, snapshotting, indexing, and live code updates, evidencing the importance of a general-purpose and self-contained design.

\section{CCF Overview}\label{sec:overview}

\begin{figure}
	\centering
	\includegraphics[scale=0.5]{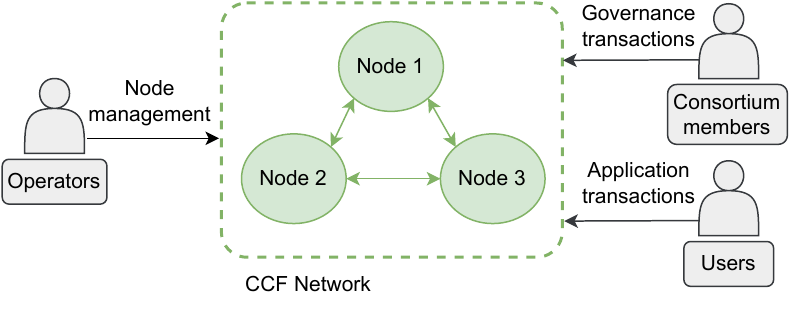}
	\caption{Overview of a CCF service. Users, operators \& consortium members interact with the three CCF nodes.}\label{fig:components}
	\Description{An example overview of a CCF service. Users, operators \& consortium members interact with the three CCF nodes via application transactions, node management, and governance transactions.}
\end{figure}

\begin{figure}
	\includegraphics[scale=0.55]{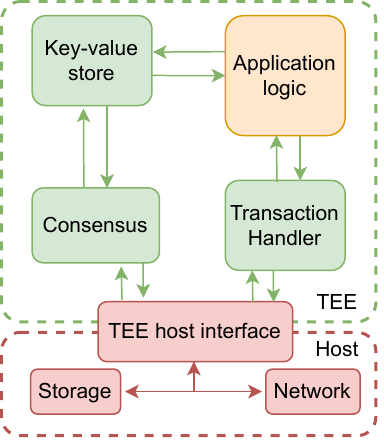}
	\caption{Key components of each CCF node.}\label{fig:node}
	\Description{The components of a CCF node, divided into trusted (TEE) and untrusted (host).}
\end{figure}

\begin{table*}\caption{Sample of the cryptographic keys utilized by nodes in CCF.}\label{fig:keys}
	\begin{tabular}{p{2cm} p{5.5cm} p{9cm}}
		\toprule
		Key (Type)                                                                                                                                      & Purpose & Lifetime \& Storage                               \\
		\midrule
		Service Identity (Asymmetric)                                                                                                                   &
		Public-key certificate (X.509) used for TLS server authentication (\autoref{subsec:users}) \& receipt verification (\autoref{subsec:receipts}). &
		Generated when a CCF service is started, either for the first time or following disaster recovery (\autoref{subsec:dr}), the associated private key is shared with all trusted CCF nodes and kept only in enclave memory. \\
		Node Identities (Asymmetric)                                                                                                                    &
		Public-key certificates (X.509) used for node authentication.                                                                                   &
		Generated separately by each CCF node when it joins the CCF service, the associated private key is stored in enclave memory and never shared.                                                                             \\
		Ledger Secret (Symmetric)                                                                                                                       &
		Used to encrypt updates to private maps in the ledger.                                                                                          &
		Shared between all trusted CCF nodes, the encrypted ledger secret is recorded in the underlying key-value store (\autoref{subsec:dr}).                                                                                    \\
		\bottomrule
	\end{tabular}
\end{table*}

A CCF service, as shown in \autoref{fig:components}, consists of a collection of CCF nodes. These nodes are deployed by one or more operators, such as cloud providers, and managed by the members of a consortium. Users invoke endpoints provided by the application logic on the CCF nodes. The application logic executes the operation corresponding to the endpoint, updating the state of the service, and responds to the user.

Endpoints vary depending on the application, for instance, a distributed logging application might simply provide \texttt{write\_message} and \texttt{read\_message} endpoints, which take a message ID and take/return a corresponding message.

On the other hand, consider a banking application, managed by a consortium of financial institutions. Endpoints such as \texttt{credit}, \texttt{debit}, and \texttt{transfer}, could take an account ID (or IDs) and an amount in USD and return with success (or an error such as insufficient funds), and the new account balance. Further endpoints might include \texttt{apply\_interest}, which updates all account balances from a given bank accordingly, or \texttt{audit}, which is available only to a financial regulator, and returns the names of account holders whose total funds exceed some threshold. 

The application-specific logic is responsible only for implementing the application itself, naming the endpoints which can be invoked and how they transform a user request to interact with the service state. CCF implements all else necessary to deploy a confidential, integrity-protected, and highly available stateful service. The programming model of CCF aims to be as easy to use as today's functions-as-a-service platforms whilst providing stronger guarantees.

CCF utilizes hardware-protected trusted execution environments (TEEs), also known as enclaves. Whilst offerings vary, in general, TEEs provide a secure and isolated execution space that guarantees that code and data are protected with respect to confidentiality and integrity. In this paper, the TEE we will be focusing on is Intel SGX (refer to \autoref{sec:related} for a discussion of other TEE platforms).
Each CCF node has its own TEE. The framework code running on CCF nodes is divided between the TEE and the untrusted host. The application logic is always executed inside the TEE. This forms the basis of CCF's confidentiality and integrity guarantees.

The application logic accesses data via a key-value store interface provided by CCF. Each endpoint invocation executes transactionally over the key-value store, potentially reading and writing from multiple keys. CCF records the result of these transactions in an append-only ledger, replicated to the other CCF nodes and to persistent storage, forming the basis of CCF's high availability guarantees.

\emph{\textbf{Threat Model.}}
Regarding safety properties such as data confidentiality, integrity, and correctness, we assume that the hosts running the CCF nodes, and by extension their operators, cannot be trusted. We also do not trust users. Attestation refers to the process by which a host can produce a verifiable proof that it has a TEE and of what code is running inside the TEE. We assume that once their attestation has been verified, the TEEs on CCF nodes can be trusted. Verification of an attestation requires trusting the hardware and the hardware manufacturer.
We assume that the code running inside the TEEs, CCF itself and the application logic, has been implemented correctly. We do not trust other applications running on the same host, the operating systems of either the guest or host, the hypervisor, or BIOS. The persistent storage is outside the trust boundary and thus could be modified or rolled back by a malicious host.

Provided that the consortium consists of multiple independent parties, together with a suitable constitution (defined in~\autoref{subsec:constitution}), then we do not trust individual consortium members. Whilst we assume that CCF nodes have TEEs, we do not assume that consortium members or users have their own TEEs.

We make the standard cryptographic assumptions such as the existence of collision-resistant hash functions. \autoref{fig:keys} gives an overview of CCF's main cryptographic keys. CCF also provides rekeying and certificate expiry/renewal, though the implementation details~\cite{CCFdocs} are beyond the scope of this paper.
In the case of catastrophic failure, CCF's disaster recovery procedure (described in~\autoref{subsec:dr}) can perform a best-effort recovery from a single copy of the ledger in persistent storage.
This recovered service presents a changed service identity, so the recovery is detectable by users (\autoref{fig:keys}).

Regarding liveness, we assume that the operators are not deliberately attacking the availability (or performance) of CCF. However, the CCF nodes are subject to the usual failures such as crash faults and network delays. We do not describe any specific measures to prevent denial-of-service attacks, but standard techniques can be applied.

\section{Intra-node Architecture}\label{sec:intranode}

In this section we examine the architecture of a static single-node CCF network; we leave the multi-node architecture (\autoref{sec:internode}) and governance (\autoref{sec:governance}) to future sections. The key components of a CCF node are shown in \autoref{fig:node}.

\subsection{Application logic}\label{subsec:users}

Users connect to the CCF node over TLS, with the connection terminating inside the TEE and the CCF \emph{service certificate} as a root of trust. Users may then invoke endpoints, provided by the application logic, using the HTTP REST API. Each endpoint invocation is executed against the latest version of the key-value store. Once completed, a transaction comprised of the request's write-set is appended to the end of the ledger and the user receives a response to their request. This response includes the ID assigned to the transaction. This \emph{transaction ID} is an ordered pair of two natural numbers, the view and the sequence number (the index of the transaction in the ledger). We will discuss views later (\autoref{sec:internode}) but for now we simply assume that transaction IDs are unique to each transaction.

Each CCF endpoint declares how callers should be authenticated. Each invocation is first checked by CCF against these declared policies and the application logic is only called if the caller passes the checks. The application logic can then implement its own authorization logic based on these authenticated claims.

\subsection{Integrity protected ledger}\label{subsec:ledger}

\begin{figure}
	\includegraphics[scale=0.75]{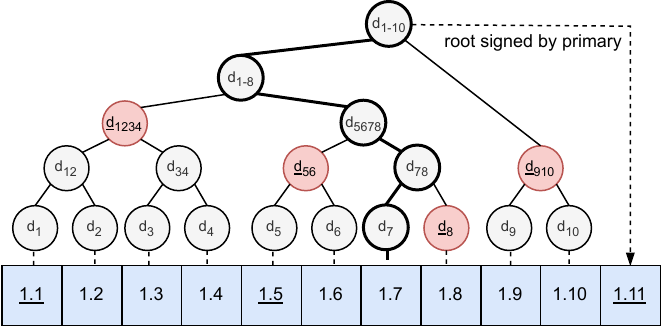}
	\caption{Example of a CCF ledger. Each box represents a transaction and the transaction ID (view \& sequence number). The underlined transaction IDs are signature transactions, which each contain a Merkle root over the ledger prefix.}\label{fig:ledger_proof}
	\Description{An example of a CCF ledger. The ledger contains 11 transactions with transaction IDs 1.1 through to 1.11. The transactions 1.1, 1.5 and 1.11 are signature transactions, which each contain a Merkle root over the ledger prefix.}
\end{figure}

As illustrated in \autoref{fig:ledger_proof}, CCF maintains a \emph{Merkle tree}~\cite{merkle87}, over the ledger. Each leaf vertex of the Merkle tree corresponds to a transaction in the ledger and each interior vertex is the digest of its two children. The Merkle tree is extended to the right as each new transaction is appended to the ledger. The root of the tree, known as the \emph{Merkle root}, is a cryptographic commitment to the whole ledger content. The CCF node periodically signs the Merkle root, and appends this to the ledger in the form of a \emph{signature transaction}. The signature transactions provide integrity protection and thus a transaction can only be considered committed once it has been followed by a signature transaction. The logical ledger is then divided into multiple physical ledger files, each terminating with a signature transaction, as it is written to persistent storage by the host.

\begin{figure}
	\includegraphics[scale=0.6]{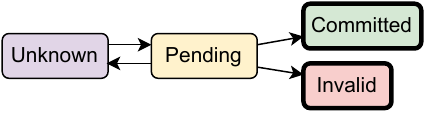}
	\caption{Transaction statuses and transitions. \texttt{Committed} and \texttt{Invalid} states are final, once observed the alternative final status will never be observed (except after disaster recovery).}\label{fig:txn_status}
	\Description{Figure shows the four possible transaction statuses, Unknown, Pending, Committed and Invalid. The transitions between these states are shown as arrows.}
\end{figure}

In addition to the endpoints exposed by the application logic, CCF also provides its own built-in endpoints common to every service. One such endpoint, \texttt{tx}, returns the transaction status (see \autoref{fig:txn_status}) of a transaction given its transaction ID. Recall that CCF replies to the user as soon as its request has finished executing and includes the associated transaction ID in the response. This \texttt{tx} endpoint thus provides one mechanism for the user to check whether the resulting transaction has been committed to the ledger.

\subsection{Transactional key-value store}\label{subsec:maps}

The underlying key-value store in CCF consists of a set of maps, each of which is a named collection of key-value pairs of a given type. Each map may either be private, meaning its updates are encrypted before leaving the TEE and being appended to the ledger, or public, in which case their updates are written to the ledger in plain text. Each transaction in the ledger includes a set of updates, each either a write-to or a removal-of a single key, to be applied atomically to the maps. These updates are subdivided into updates to public maps (unencrypted) and updates to private maps (encrypted).

The current state of the maps is kept in memory on the CCF node and the key-value store is utilized both by the application logic and by CCF itself. For instance, the map \texttt{ccf.internal.\allowbreak{}signatures} stores the signed Merkle root from the signature transaction. CCF's internal and governance maps are public, enabling auditing without ledger decryption.

\subsection{Read-only \& historical transactions}\label{subsec:reads}

Not all transactions update the key-value store, and as such CCF provides a fast path for read-only transactions. Read-only transactions are executed against the latest version of the key-value store but are not then recorded on the ledger. The user receives a response from CCF as per usual, except the transaction ID included is the ID of the last transaction applied to the key-value store instead of a new one.

The application logic can also choose to query the historical state of the ledger over a range of sequence numbers. Since historical queries may require fetching and processing many transactions from persistent storage, they can take some time. To accelerate historical range queries, CCF enables applications to define their own indexing strategy within the application logic. The indexer on the CCF node pre-processes in-order each transaction in the ledger as it is committed and stores the results for future use. Alternatively, this can also be done lazily when a historical query is received. The indexer's state is kept in-memory for quick access but can be offloaded to persistent storage if needed.

As an example, an application's indexing strategy could record for each key every transaction ID that writes to the key. Such an index could be utilized by our banking example from earlier to implement a \texttt{get\_statement} endpoint that returns all recent credits/debits given an account ID.

\subsection{Verifiable receipts}\label{subsec:receipts}

\emph{Receipts} are signed statements, associated with a given transaction, that can be verified and audited offline. Like the key-value store, receipts are utilized both internally by CCF, to ensure the validity of snapshots (\autoref{subsec:reconfiguration}), and externally by users. Receipts can be leveraged either to audit CCF or to prove to a third party that a transaction was committed in a particular context and at a specified position in the ledger. One way users can generate a receipt is by invoking the built-in \texttt{receipt} endpoint with the transaction ID.

At its core, the receipt includes a signature over the Merkle root (from a subsequent signature transaction) and the sequence of vertices on the path from the transaction leaf to the root, known as a \emph{Merkle proof}. In \autoref{fig:ledger_proof}, the Merkle proof for transaction 1.7 is $[(right, d_8),(left,d_{56}),(left, d_{1234}),(right, d_{910})]$. The Merkle proof can be used to verify that the transaction leaf, which includes the transaction ID and a digest of the write-set, was in the ledger at the time of the signature.
The application logic may also choose to attach arbitrary claims to a transaction and thus its receipt to make them verifiable offline.
\section{Inter-node Architecture}\label{sec:internode}

Last section we focused on a single-node CCF network.
This section describes how we extend that to a multi-node network, providing high availability in the presence of crash faults.
Moreover, multi-node CCF networks also increase the throughput of read transactions, which can be handled by any node.
CCF includes its own consensus layer for replicating transactions in the ledger, inspired by Raft~\cite{Ongaro14} and adapted for a fault model more suited to trusted execution.
This protocol requires $n$ CCF nodes to tolerate $f=\lfloor \frac{n-1}{2} \rfloor$ crash faults.
To build confidence in our approach, we used TLA+~\cite{SpecifyingSystems} to formally specify and model check CCF's consensus layer including reconfiguration~\cite{CCF-TLA}.

\subsection{Transaction replication}\label{subsec:replication}

\begin{figure}
	\includegraphics[scale=0.45]{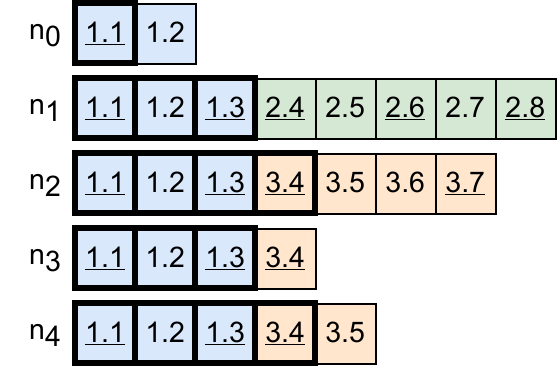}
	\hspace{0.1cm}
	\includegraphics[scale=0.45]{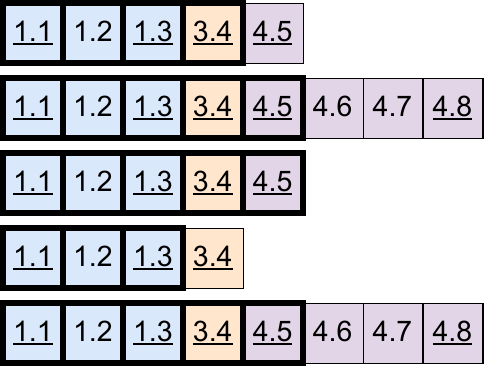}
	\caption{Examples of five CCF ledgers. Underlined transactions IDs are signature transactions and the thick boxes highlight the committed transactions.}\label{fig:election_ledgers}
	\Description{Figure shows five CCF ledgers at two different times.}
\end{figure}

\begin{table}
	\caption{Possible votes and primaries during an election, based on the ledgers shown in \autoref{fig:election_ledgers} (left). For each possible candidate (row), the table shows which nodes might vote for it and thus whether it could win the election.}\label{tab:election_votes}
	\begin{tabular}{ccccccc}
		\toprule
		candidate & $n_0$    & $n_1$    & $n_2$    & $n_3$    & $n_4$    & could win? \\
		\midrule
		$n_0$     & \cmark{} & \xmark{} & \xmark{} & \xmark{} & \xmark{} & \xmark{}   \\
		$n_1$     & \cmark{} & \cmark{} & \xmark{} & \xmark{} & \xmark{} & \xmark{}   \\
		$n_2$     & \cmark{} & \cmark{} & \cmark{} & \cmark{} & \cmark{} & \cmark{}   \\
		$n_3$     & \cmark{} & \cmark{} & \xmark{} & \cmark{} & \cmark{} & \cmark{}   \\
		$n_4$     & \cmark{} & \cmark{} & \xmark{} & \cmark{} & \cmark{} & \cmark{}   \\
		\bottomrule
	\end{tabular}
\end{table}

Time is divided into a series of \emph{views}. During most views, one of the nodes is the \emph{primary} and the remaining nodes are the \emph{backups}. The primary executes the user requests, appends the resulting transaction to its ledger, and replies to the users. The primary regularly replicates its ledger onto the other nodes using the integrity-protected \texttt{append\_entries} remote procedure call (RPC).

As before, the primary regularly appends signature transactions. The primary tracks the matched sequence number on each node and once a signature transaction has been copied onto $\lceil \frac{n-1}{2} \rceil$ other nodes then it considers the signature transaction and all previous transactions as committed and updates its \emph{commit sequence number} accordingly. This updated commit sequence number is included in future \texttt{append\_entries} so that the backups can update their local commit sequence number.

Recall that each transaction ID uniquely identifies a transaction. Each \texttt{append\_entries} includes a set of transactions and the previous transaction ID. The backup checks that the previous transaction ID matches its ledger before appending the new transaction. This check ensures that if any two ledgers contain a transaction with the same ID then the ledgers up to and including that transaction are identical.
\autoref{fig:election_ledgers} shows an example of the ledgers from five nodes, where node $n_2$ is the primary in view 3. Note that the last committed transaction is always a signature transaction and that all committed transactions are present in at least a majority of ledgers.

\subsection{Primary elections}\label{subsec:view_change}

\begin{figure}
	\includegraphics[scale=0.6]{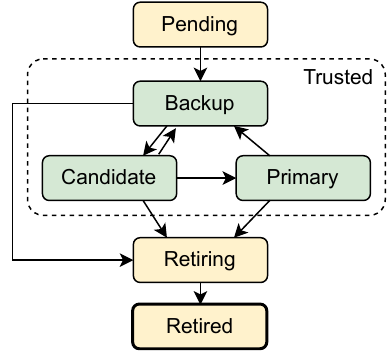}
	\caption{Node states \& transitions: \texttt{PENDING}, \texttt{TRUSTED} including \texttt{PRIMARY}, \texttt{CANDIDATE}, \& \texttt{BACKUP}, \texttt{RETIRING}, and \texttt{RETIRED}.}\label{fig:states}
	\Description{Figure shows the six possible node states and the transactions between them. A pending node can transition to a backup, a backup can transition to candidate. A candidate can transition to primary or back to a backup and a primary can transition to a backup. Any trusted node (primary, candidate, or backup) can transition to retiring and a retiring node can transition to a retired node.}
\end{figure}

Eventually, one or more nodes will fail. If the failed node (or nodes) are backups, and $\lceil \frac{n-1}{2} \rceil$ backups remain then the primary will continue as before. If the primary fails then a new primary will be elected, as illustrated by \autoref{fig:states}.

Note that each node keeps track of its perspective on the current view, increasing it accordingly as messages are exchanged. Each RPC includes the sender's view, and the receiver will update its view if the sender's view is greater before processing the message. If the sender's view is less than the receiver's view then the receiver will not process the message. Instead, it will reply negatively to the message including its view so that the sender can update its own view. When a node increases its view, it becomes a backup if it is not already one.
The primary sends regular heartbeats, in the form of empty \texttt{append\_entries} RPCs to its backups. If a backup does not hear from a primary within its election timeout, it becomes a \emph{candidate}, increments its view, and dispatches a \texttt{request\_vote} RPC to all nodes including the view and sequence number of its last signature transaction. After checking the candidate's view, the recipient of the \texttt{request\_vote} RPC checks if the candidate's ledger is at least as up-to-date as its own, and if so, provided it has not voted for another candidate this view, it will vote for the candidate. In doing so, the backup also resets its own election timeout.

A candidate's ledger is at least as up-to-date as the voter's ledger if
\begin{enumerate*}[label=(\roman*)]
	\item the view of the candidate's last signature transaction is greater than the view of the voter's last signature transaction, or
	\item the views are equal and the sequence number of the candidate's last signature transaction is greater or equal to the sequence number of the voter's last signature transaction.
\end{enumerate*}

An example of this voting criteria is shown in \autoref{tab:election_votes}. $n_0$ cannot become primary as the other nodes would not vote for it. $n_1$ cannot become primary as only one other node ($n_0$) would vote for it and $n_2$ could become primary as all nodes would vote for it. $n_3$ and $n_4$ could become primary as $n_0$ and $n_1$ could vote for them, and they could vote for each other.

If a candidate receives votes from $\lceil \frac{n-1}{2} \rceil$ other nodes, then it becomes a primary. It discards any transactions after the last signature transaction and begins sending \texttt{append\_entries} to its backups. The new view will begin with a signature transaction and the primary can update its commit sequence number to this transaction as soon as it has been replicated to the majority of ledgers. The new primary tracks the sequence number of the next transaction to send to each backup, starting with the first signature transaction of the next view.
The ledgers of the backups might have fallen behind or even diverged from that of the new primary. Recall that when a backup receives an \texttt{append\_entries} it checks that it has the previous transaction ID. If not, it responds negatively to the primary including the sequence number of what it believes to be the latest common point. The primary decrements its sequence number for that primary, utilizing the information provided by the backup, and tries again until the \texttt{append\_entries} is successful. If a backup has conflicting transactions then they are deleted before the new transactions are appended. In other words, the primary's ledger is considered the ground truth when there are conflicts.

Consider the example on the right of \autoref{fig:election_ledgers}. Following the situation on the left of \autoref{fig:election_ledgers}, $n_4$ became the primary in view four. After rolling back transaction 3.5, $n_4$ added signature transaction 4.5 and appended it to $n_0$, $n_1$ \& $n_2$. $n_4$ also appended transactions 4.6, 4.7, and signature transaction 4.8 to its ledger but has so far only replicated these to $n_1$.

If a candidate's election timer expires without it either stepping up to a primary or down to a backup then it restarts the election by incrementing its view once more and sending new \texttt{request\_vote} RPCs.
It might be the case that a new primary is elected whilst the previous primary is still alive. In this scenario, the new primary will eventually learn of the new view and will step down to a backup in the process. Multiple backups can become candidates at a similar time, splitting the votes. To reduce the likelihood of this, the election timer is randomly chosen from a configurable range.
The primary also keeps track of the last time it received an \texttt{append\_entries} response from each backup, and it steps down if it does not hear from at least $\lceil \frac{n-1}{2} \rceil$ backups within a specified time window. This ensures that a primary that cannot make progress, for instance due to a partial network partition, steps down cleanly, allowing another node to become primary, and does not produce an arbitrarily long ledger suffix which will never be committed.

\subsection{User interaction}\label{subsec:forwarding}

Users can connect to any CCF node, and backup nodes can forward user requests to the primary. When a node fails, the users connected to it will retry with other nodes. Read-only requests, including receipt requests, can be served by any CCF node and do not need to be forwarded to the primary. However, to ensure that CCF provides session consistency to its users, once a backup has forwarded a request to a primary it will forward all subsequent requests on the same TLS session, including reads, to the same primary, and terminate the session if this forwarding is not possible due to a primary change.

Note that each node maintains a view history, containing the start index for each view, according to its ledger. This history can help a node determine transaction status (\autoref{fig:txn_status}), for instance, a transaction has status \texttt{Invalid} if a greater view started at a smaller sequence number.

\subsection{Atomic reconfiguration}\label{subsec:reconfiguration}

When a node fails, the CCF network can tolerate one fewer fault than it could before. To restore the fault tolerance of the CCF network, the failed node should be replaced with a new node. This process, known as \emph{reconfiguration}, allows an arbitrary transition from any node configuration to any other node configuration. For instance, any number of nodes can be added, removed, or a combination of both, atomically changing the total number of nodes and thus changing the fault tolerance accordingly.

Reconfiguration is achieved by a reconfiguration transaction. Reconfiguration transactions are similar to signature transactions, in that they contain updates to built-in maps and affect the operation of a CCF node directly, distinguishing them from user-defined updates to application tables. For reference, \autoref{tab:ccf_maps} provides a sample of some built-in maps used by CCF.

\begin{table}\caption{Sample of CCF's built-in maps. Some names have been edited for brevity and all are prefixed with \texttt{public:ccf.gov} or \texttt{public:ccf.internal}.}\label{tab:ccf_maps}
	\begin{tabular}{l l}
		\toprule
		Name                                  & Purpose                                               \\
		\midrule
		\texttt{users.certs}                  & User certificates                                     \\
		\texttt{members.certs}                & Consortium member certificates                        \\
		\texttt{nodes.info}                   & Node identity certificates \& properties              \\
		\texttt{nodes.code\_ids}              & Code versions that are allowed to join                \\
		\texttt{service.info}                 & Service identity certificate \& service status        \\
		\texttt{signatures}                   & Merkle roots and signatures                           \\
		\texttt{tree}                         & Merkle tree for historical receipts                   \\
		\texttt{history}                      & Governance operations signed by members               \\
		\texttt{constitution}                 & CCF service constitution                              \\
		\texttt{modules}                      & JavaScript application logic                          \\
		\texttt{endpoints}                    & JavaScript endpoints                                  \\
		\texttt{proposals}                    & Governance proposals                                  \\
		\texttt{proposals\_info}              & Status of governance proposals \& ballots             \\
		\texttt{ledger\_secret}   & Encrypted ledger secret                   \\
		\texttt{recovery\_shares} & Encrypted shares to recover ledger secret \\
		\texttt{members\_keys}    & Members' public encryption keys           \\
		\bottomrule
	\end{tabular}
\end{table}

A reconfiguration transaction includes writes to \texttt{nodes.info} which (among other things) maps each node's ID to a node status (\autoref{fig:states}). The current configuration is the set of nodes whose status is \texttt{TRUSTED}. Reconfiguration transactions can add a node by changing its status from \texttt{PENDING} to \texttt{TRUSTED} and remove a node by changing its status to \texttt{RETIRED}. The question of how nodes are first added to the map as \texttt{PENDING} and removed from the map is addressed in \autoref{sec:governance}.

Each node maintains a sorted list of \emph{active configurations}, where each configuration contains the transaction ID of the reconfiguration transaction and the associated set of nodes. At the head of the list is the current configuration, followed by any pending configurations. Winning an election or committing a transaction requires a majority quorum in each active configuration, therefore \texttt{request\_vote} and \texttt{append\_entries} are sent by the candidate or primary to all nodes across the active configurations. As soon as a node appends a reconfiguration transaction to its ledger, regardless of whether the node is a primary or backup, that configuration is added to the active configurations. The reconfiguration transaction, like user transactions, is considered committed by the primary only when a subsequent signature transaction has been committed on all active configurations. When a node learns that a reconfiguration transaction has been committed, all earlier configurations are removed from the active configurations. A newly added node begins participating in consensus, including potentially starting an election, as soon as it appends to its ledger the first signature transaction following the reconfiguration transaction that added it.

The process described thus far requires new nodes to replay all transactions in the history of the CCF service in order to become up to date. In practice, nodes can begin from a snapshot and use the consensus layer to simply learn the transactions since the last snapshot.

Recall that nodes add new configurations to the set of active configurations as soon as the associated reconfiguration transaction is appended to their ledger, they do not wait for the reconfiguration transaction to be committed or even followed by a signature transaction. Uncommitted transactions however can be rolled back following a view change. When this occurs, the rolled-back reconfigurations are removed from the set of active configurations. Note that the set of active configurations will always contain at least one configuration, the current configuration, which is either the initial configuration or the last committed reconfiguration, which cannot be rolled back.

\subsection{Node retirement}

The process of retiring nodes includes an additional step. The reconfiguration transaction first updates the node's status to \texttt{RETIRING}. As soon as the reconfiguration transaction is committed, the previous configurations are removed as before and the retired node is no longer part of the configuration. Before we shut down the retiring node, the primary must ensure that all future primaries also know that the node has been successfully retired. Therefore, once this reconfiguration transaction has been committed, the primary adds a second reconfiguration transaction to update the node's state to \texttt{RETIRED}. Once this second transaction is committed, then the retired node can be shut down. If there is a view change during reconfiguration, the new primary must commit the first reconfiguration transaction before appending the second reconfiguration transaction.

A primary may commit a reconfiguration transaction that retires itself. In this case, it should stop adding new transactions after the signature transaction which commits its retirement. Once the first reconfiguration transaction is committed, it should then stop sending heartbeats but remain online, replicating its ledger and voting for new primaries, whilst a new primary establishes itself. As before, the retiring node can be shut down once the new primary commits the reconfiguration transaction which sets its state to \texttt{RETIRED}.

\section{Multiparty Governance}\label{sec:governance}

CCF is designed to support long-running applications and thus must remain available in the face of typical management operations such as node replacement, live code updates, changes to the set of users, and more. Such actions must be done with great care to preserve CCF's strong confidentiality and integrity properties. Moreover, since CCF services can be managed by multiple parties, these parties should have an opportunity to sign off on management operations before they are executed.

\subsection{Governance proposals \& ballots}\label{subsec:constitution}

At a high level, CCF's approach to multiparty governance is to require consortium members to oversee governance operations using governance proposals and ballots, which are processed by the programmable constitution.

The \emph{constitution} is a contract between the consortium members describing all the available governance actions and the associated voting criteria. The constitution, encoded in JavaScript, is provided to a CCF service at start-up. The constitution defines a \texttt{resolve} function, which takes a governance proposal (defined below) and votes by consortium members, and determines if the proposal has been accepted. The constitution also defines \texttt{apply}, which takes an accepted proposal and executes the governance actions within it to modify the key-value store. The default constitution~\cite{CCF-cons} accepts a governance proposal once it has received votes from a strict majority of consortium members. Alternative constitutions could give consortium members different voting power, for instance giving a few consortium members, who are the main parties running the application, veto power. Alternatively, a constitution could vary the number of votes needed depending on the action, for instance giving one consortium member, who is also an operator, the unilateral power to add or remove nodes. The constitution can also be updated using a governance proposal (if permitted by the current constitution).

A \emph{governance proposal} consists of a set of actions, defined in the constitution, and a \emph{ballot} is a vote for or against a particular proposal, conditional on the proposal itself and the current state of the key-value store. Similar to user transactions, governance proposals and ballots are recorded on the underlying CCF ledger, However, unlike user transactions, they always originate from a request signed by a consortium member. This signature is also stored on the ledger.

\begin{table}\caption{Sample of the governance actions defined in CCF's default constitution~\cite{CCF-cons}.}\label{tab:gov_actions}
	\begin{tabular}{ll}
		\toprule
		Governance action                             & Purpose                               \\
		\midrule
		\texttt{set\_user}                            & Add user                              \\
		\texttt{set\_member}                          & Add consortium member                 \\
		\texttt{set\_js\_app}                         & Update JS application logic           \\
		\texttt{add\_node\_code}                      & Allow a new code ID                   \\
		\texttt{transition\_node\_to\_trusted}        & Allow node to join                    \\
		\texttt{set\_constitution}                    & Update constitution                   \\
		\texttt{transition\_service\_to\_open}        & Allow user requests                   \\
		\texttt{set\_recovery\_threshold} & Update recovery threshold \\
		\bottomrule
	\end{tabular}
\end{table}

\begin{lstfloat}
	\begin{lstlisting}[frame=single,style=ES6,basicstyle=\ttfamily\scriptsize]
["add_node_code", new Action(
  function (args) { 
    checkType(args.code_id, "string", "code_id") },
  function (args, proposalId) {
    ccf.kv["public:ccf.gov.nodes.code_ids"].set(args.code_id, "AllowedToJoin");
    invalidateOtherOpenProposals(proposalId)})]
\end{lstlisting}
	\caption{An example action from the \texttt{apply} function in CCF's default constitution. Adapted for brevity.}\label{lst:node_code}
\end{lstfloat}

\autoref{tab:gov_actions} shows a sample of the governance actions defined in the \texttt{apply} function of the default constitution and \autoref{lst:node_code} shows the implementation of one such action. This action, \texttt{add\_node\_code}, can be used when updating the code inside the TEE (CCF itself or its C++ application logic), and adds the given code ID to the set of trusted code IDs. This is done by writing the code ID to the \texttt{nodes.code\_ids} map (listed in \autoref{tab:ccf_maps}) in the key-value store. That map is then read from when a new node attempts to join the service, to establish whether they are running a trusted version of the code.

Governance proposals have the power to change many fundamental properties of the service, not least modifying the constitution itself. Any errors, deliberate or otherwise, are therefore potentially damaging. Consortium members should examine proposals carefully before submitting their ballot to ensure they do not leak confidential state. Proposals are encoded as succinct JSON documents so that they are easy to inspect offline. Note that the default constitution~\cite{CCF-cons} provided by CCF should be suitable for most deployments.

\subsection{Disaster recovery protocol}\label{subsec:dr}

Governance and reconfiguration require consensus and thus majority agreement. If more than a majority of nodes fail simultaneously then CCF cannot make progress. In this case, the disaster recovery protocol can enable the service to restart. Note that disaster recovery restores the CCF service to a safe state, even from just a single copy of the ledger files, but it is not guaranteed to preserve all committed transactions. This best-effort approach is necessary when any consensus-based service wishes to recover from the persistent storage of just a minority of nodes.
The newly recovered service will have a new service identity (defined in \autoref{fig:keys}), making it clear to users that a disaster recovery has occurred, and thus that the service may have been rolled back to a previous state.

During disaster recovery, a node or set of nodes is started in recovery mode with a set of ledger files from the previous CCF service (and optionally, snapshot files to speed things up). The public parts of transactions are restored, and further nodes may join the network. A consortium member will submit a proposal to open the new service, specifying the old and new service identities to ensure this proposal is bound to a precise recovery, and once the proposal receives sufficient approving ballots, the service is opened.

In order to decrypt the updates to private maps, consortium members must submit their recovery shares to the CCF service. Recall that updates to private maps are encrypted using the ledger secret, as defined in \autoref{fig:keys}. The ledger secret is recorded in the ledger, encrypted using the \emph{ledger secret wrapping key}. The ledger secret wrapping key is itself stored in the ledger after being split into $n$ shares, known as \emph{recovery shares}, where $n$ can be up to the number of members in the consortium. Each recovery share is first encrypted using one of the consortium member's public encryption keys. The ledger secret wrapping key is divided into $k$-of-$n$ shares~\cite{Shamir79}, such that $n$ recovery shares are created, less than $k$ shares leak nothing about the ledger secret, and any $k$ shares suffice to recover the ledger secret. Once these shares are submitted by the consortium members, the ledger secret can be reconstructed in TEE, and the previous ledger's private state decrypted. Once this is complete, the CCF service is fully recovered. The parameter $k$, known as the recovery threshold, can be configured through governance. \autoref{tab:ccf_maps} includes a summary of the built-in maps utilized by CCF's disaster recovery procedure.
\section{Discussion}\label{sec:discussion}

We set out to investigate whether it is possible to develop a pragmatic solution for deploying confidential, integrity protected, and highly available multiparty applications on untrusted cloud infrastructure. We now evaluate CCF along these dimensions.

\subsection{Confidentiality}\label{subsec:confidentiality}

Keeping data confidential requires protection during every stage of a CCF service.
Firstly, data is encrypted in transit between users and CCF nodes with a TLS connection, using the service identity certificate (defined in \autoref{fig:keys}), and terminating inside the TEE. Data is encrypted during execution and in memory by the TEEs. Data is encrypted at rest (on disk) and in transit between CCF nodes as the transactions in the ledger are encrypted with the ledger secret (\autoref{fig:keys}). The service certificate private key and ledger secret are not shared with new CCF nodes until their attestation has been verified.
Whilst the application logic has read-write access to the key-value store (except CCF's read-only internal/governance maps), the application logic can limit which users can invoke specific endpoints using user authentication policies. The application logic and set of authorized users can only be modified via a governance proposal.
There is a fundamental tradeoff between confidentiality and auditability; some transactions, including all governance operations, are written unencrypted to permit offline audit, while application data which does not require such audit can be encrypted by the ledger secret.

\subsection{Integrity}\label{subsec:integrity}

Whilst the ledger secret is the primary mechanism for providing confidentiality, it is the signature transactions (\autoref{subsec:ledger}) that provide integrity protection to the ledger whilst outside the TEEs. The Merkle tree constructed for the signatures transactions is again utilized to provide Merkle proofs for receipts (\autoref{subsec:receipts}).
Integrity protection with signature transactions ensures that a malicious party cannot modify the ledger undetected whilst it is in persistent storage, however, the ledger could be rolled back to a previously valid prefix.
Notably, CCF's design avoids the need to depend upon a dedicated rollback protection mechanism. Existing rollback protection approaches include SGX trusted monotonic counters~\cite{TrustedTime17} (which have since been deprecated~\cite{NoMoreMono}), using TPMs~\cite{NoMoreMono,Martin21}, or a service based on replication across TEEs such as ROTE~\cite{Matetic17}, LCM~\cite{Brandenburger17}, Tiks in Engraft~\cite{Wang22}, NARRATOR~\cite{Niu22} or Nimble~\cite{Angel23}. Rollback protection is also closely related to forking protection mechanisms and likewise, CCF avoids the need for them by design. The architecture of CCF treats nodes as ephemeral. When a node crashes, it cannot simply resume from persistent storage as we cannot be sure that its state is up-to-date. Instead, the CCF node rejoins the service via reconfiguration (\autoref{subsec:reconfiguration}) and establishes a fresh node identity. This stricter fault model as well as the need for integrity protection are the motivation behind CCF's choice of a custom consensus protocol over a classic approach such as Multi-Paxos~\cite{Lamport98} or Raft~\cite{Ongaro14}.

CCF adopts a transparent approach to management where governance proposals and ballots are recorded in public maps and signed by consortium members to enable auditability and disaster recovery (\autoref{subsec:constitution}).

\subsection{High availability}\label{subsec:availability}

CCF achieves high availability through replication across CCF nodes (\autoref{sec:internode}). The consensus layer in CCF uses strict majority quorums thus it can continue to operate even when up to a minority of nodes fail simultaneously. Users can simply connect to any CCF node, requests are handled locally if they are read-only, including historical transactions and receipt generation, and forwarded to the primary otherwise (\autoref{subsec:forwarding}). If the CCF node that a user is connected to fails, then the user can retry with another node.

Failed nodes can be replaced by reconfiguration to restore the fault tolerance of the service (\autoref{subsec:reconfiguration}). For instance, a CCF service with five nodes can tolerate two faults, if one node fails then the service can only tolerate one further fault. Reconfiguring to remove the failed node and add a new healthy node restores the fault tolerance of the service back to two faults. In addition to recovering from failures, reconfiguration can also be used proactively to migrate the service, to change the replication factor, or for code updates, either to CCF itself or to the application logic. CCF's consensus layer supports arbitrary reconfiguration, moving from any set of nodes to any other set of nodes, without stopping the service. CCF also achieves reconfiguration in just a single transaction, with a second transaction only needed to confirm node removal.
The classic approaches to reconfiguration in Raft either requires nodes to be added or removed one at a time~\cite{etcdReconfig,diego}, or always requires two transactions~\cite{Ongaro14}.
Our design is particularly useful as it enables CCF services to minimize the window of vulnerability until the usual degree of resilience is restored.
If more than a majority of nodes are lost, CCF can recover from persistent storage through its disaster recovery procedure (\autoref{subsec:dr}).	During disaster recovery, the service may be reverted to a prior state (a suffix of the ledger may be lost).

The constitution (\autoref{subsec:constitution}) defines what groups of consortium members must vote in favor of a proposal for it to be accepted, for instance, a majority quorum. Votes are recorded and calculated using the ledger and thus consortium members do not need to be online simultaneously to propose and accept a proposal. Proposing and voting for a proposal can be done on the fly and does not require stopping the service.
If a sufficient number of consortium members lose their keys, then governance operations and disaster recovery will no longer be possible.

\subsection{Pragmatism}\label{subsec:pragmatic}

CCF is designed to make the deployment of trustworthy multiparty applications on untrusted infrastructure as simple as possible. CCF supports custom application logic that can be written in either JavaScript/TypeScript or C++, providing greater flexibility compared to permissioned blockchains which often support DSLs such as Solidity, Move~\cite{Move}, or Daml.
The programming model is simple, application logic exposes endpoints that are invoked by users, and state changes are recorded by a transaction over the built-in data store. Transactions are applied to the key-value store atomically and in isolation thus the application will never observe partial transaction execution. Application logic is not required to be deterministic, it may be executed multiple times but its associated transaction is applied exactly once.
The constitution is also fully programmable, allowing the consortium members granular control over governance operations. The constitution, written in JavaScript, is provided at start-up and can be updated by a governance proposal (if supported by the current constitution).

Many aspects of CIA support in CCF are configurable, allowing CCF to adapt to different deployment scenarios. Maps in the key-value store can be either private (encrypted on the ledger) or public (plain-text). Public maps are not confidential but can be audited without decrypting the ledger, and thus can be audited offline. Application logic maps are private by default, and public maps are also supported. CCF's built-in maps are public for transparency. CCF can be deployed on hosts without TEEs using \emph{virtual} mode. This is useful for development and, in practice, for replication if confidentiality and integrity are not needed. CCF in virtual mode still provides highly available state machine replication with advanced features such as atomic reconfiguration, indexing, snapshots, historical transactions, disaster recovery, receipts, and multiparty governance. Likewise, CCF can be deployed on a single node if high availability is not needed.

Users of CCF can determine on a per-request basis the consistency/durability guarantee required. CCF provides two levels of guarantee: local execution \& global commit. Users receive a reply with a transaction ID as soon as it has been executed locally. The user can determine after having observed the response if it wishes to proceed without checking for commit. Likewise, receipts can be expensive to retrieve and thus are issued on demand, at the user's request, again using the transaction ID. There is optional support for user request signing, via the same mechanism that consortium members sign governance operations. CCF also allows applications to define their own indexing strategy.

CCF currently supports Intel SGX, and it is designed with a view to supporting more TEE platforms, for instance, by minimizing dependencies on features specific to SGX (or specific versions of SGX) e.g. sealing~\cite{Sealing16}, trusted time and monotonic counters~\cite{TrustedTime17}, or ECALL/OCALL interface provided by SGX.

\section{Implementation \& Evaluation}\label{sec:evaluation}

CCF is written in C++20 and built on the OpenEnclave SDK~\cite{OE}.  The host and the TEE communicate via a pair of lock-free multi-producer single-consumer ringbuffers to minimize the expensive transitions to/from the TEE~\cite{SGXperf,Orenbach17,Arnautov16,Weisse17}. CCF uses a host-side thread, a TEE-side thread, and optionally, a TEE-side thread pool with a configurable number of worker threads. CCF's REST API supports HTTP/1.1 and HTTP/2 with a custom HTTP header in responses for the transaction ID. Built-in endpoints use JSON and COSE payloads. CCF supports TLS 1.2 \& 1.3 using OpenSSL 1.1.1~\cite{openSSL}.  Request signing by consortium members, and optionally by users, supports HTTP signatures~\cite{RFC-draft-http-signatures-12} and COSE Sign1~\cite{RFC8152}. User authentication is implemented using JWT (JSON Web Tokens)~\cite{RFC7519} and by X.509 certificates~\cite{RFC5280}.
Our Merkle tree implementation~\cite{MerkleCPP} uses SHA-256~\cite{SHA}. The ledger secret uses AES256-GCM~\cite{RFC5288} to encrypt the transactions on the ledger. AES256-GCM is also used to encrypt any persistent storage utilized by the indexer. Diffie-Hellman key exchange~\cite{Diffie76} is used for node-to-node message headers and message forwarding. The encryption of ledger secrets uses RSA OAEP~\cite{Bellare94}. The map implementation in CCF is based on the Compressed Hash-Array Mapped Prefix-tree (CHAMP)~\cite{Steindorfer18}. CCF's JavaScript engine is QuickJS~\cite{QuickJS}.

\textbf{\emph{Experiment Setup.}} Unless specified otherwise, our setup was as follows. We used DC16s\_v3 (16 cores, 128 GB memory) VMs~\cite{DCsv3} in the UK South region of Azure, running Ubuntu 20.04. Our Ice Lake CPUs support Intel SGX2~\cite{sgx2}. Our C++ application logic implements a simple logging application, where messages with corresponding identifiers are posted, and later retrieved with read-only transactions. Messages are private and 20 characters each. We use four VMs: three CCF nodes, configured with 10 worker threads, with one user VM, simulating sufficient users (closed loop with up to 1k concurrent requests) to stress test CCF. To measure the performance of CCF itself, instead of the optional node-to-node forwarding logic, the user directly writes to the primary.

\begin{figure}
	\centering
	\includegraphics[scale=0.5]{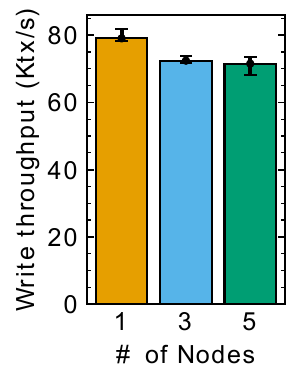}
	\includegraphics[scale=0.5]{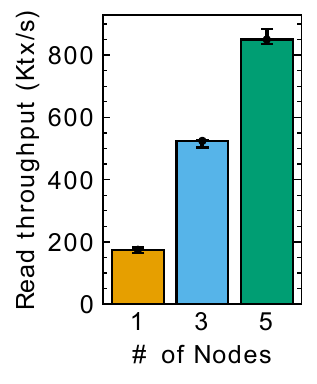}
	\includegraphics[scale=0.5]{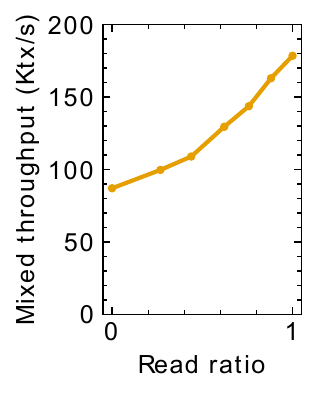}
	\caption{Impact of the number of CCF nodes on writes (left) and reads (center) and the impact of read/write request ratio on single node throughput (right).}\label{fig:throughput_scale}
\end{figure}

\autoref{fig:throughput_scale} shows that CCF consistently achieves a throughput of at least 65 thousand requests per second. Increasing the ratio of reads to writes slightly increases performance. This effect is most notable for the 100\% read workload, where all user requests are handled directly by the CCF node that receives them since only writes must be handled by the primary. We also observed similar performance using public maps instead of private ones.

\begin{table}\caption{Throughput (tx/s) for writes/reads on a five node service. Virtual mode refers to CCF running without SGX.}\label{tab:throughput}
	\begin{tabular}{l rr}
		\toprule
		    & SGX             & Virtual        \\
		\midrule
		C++ & 64.8 K / 881 K  & 118 K / 1.24 M \\
		JS  & 15.7 K / 90.7 K & 33.7 K / 219 K \\
		\bottomrule
	\end{tabular}
\end{table}

\autoref{tab:throughput} illustrates the significant performance difference between the C++ implementation of the logging application and the JavaScript implementation. \autoref{tab:throughput} also demonstrates the significant overhead of Intel SGX. We hope that future TEEs will substantially reduce this overhead such that performance is comparable to that of non-confidential offerings. This is supported by early benchmarks of AMD SEV-SNP which report a performance overhead of just 2--8\%~\cite{SNPPerfPost}.

\begin{figure}
	\includegraphics[scale=0.5]{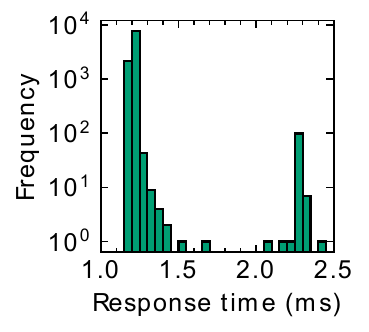}
	\includegraphics[scale=0.5]{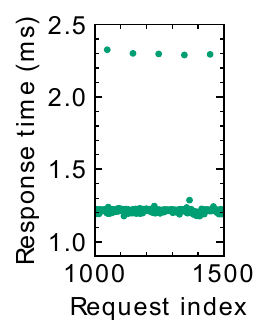}
	\includegraphics[scale=0.5]{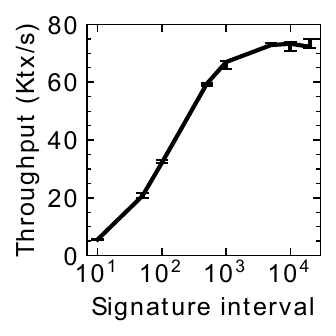}
	\caption{Impact of the signatures on response time (left \& center) and write throughput (right).}\label{fig:latency}
\end{figure}

\autoref{fig:latency} investigates the performance impact of generating signature transactions. To measure the overhead of signatures, the signature transaction frequency has been set to every 100 transactions and most other sources of latency variance removed, for instance, we use one node and one user. We then measure the \emph{response time}, the time for the user to receive a positive response from the CCF node, including a transaction ID that can then be used to check for commit. We observe that the response time of write requests is 1.2--1.3 ms, with the exception that approximately every 100 requests, a signature is triggered, and the response time increases to 2.3 ms. The graph on the right shows how the throughput of writes is impacted by the frequency of signature transactions, demonstrating the tradeoff between time to commit (which requires a signature), and overall throughput.

\begin{figure}
	\includegraphics[scale=0.5]{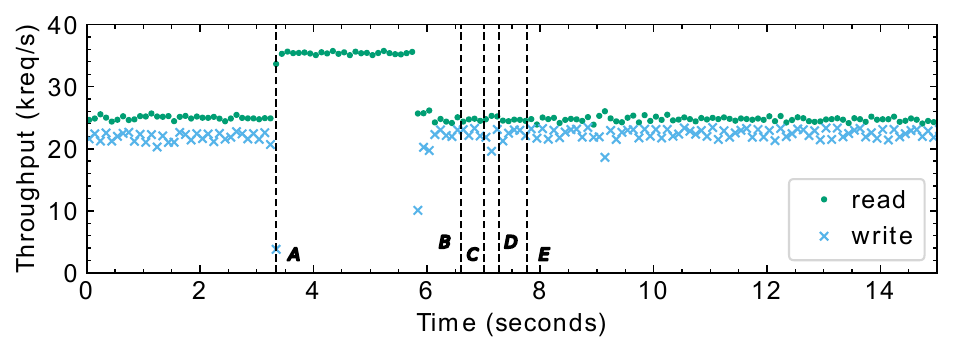}
	\caption{Impact of primary failure ($A$) and subsequent node replacement ($B$-$E$) on the availability of reads and writes.}\label{fig:availability}
\end{figure}

\autoref{fig:availability} shows the impact of a primary failure and subsequent reconfiguration on a CCF service with three initial nodes $\{n_0, n_1, n_2\}$, and three consortium members $\{m_0, m_1, m_2\}$ using the default constitution. \autoref{lst:reconf} also shows five governance transactions from the ledger in \autoref{fig:availability}. The graph shows two users, one sending writes to $n_0$, the initial primary, and the other sending reads to $n_1$, one of the two backups. At $A$, we kill $n_0$, writes stop immediately as the service is without a primary and the reads continue (in fact their throughput increases as the backup is no longer handling messages from the primary). $n_2$ is then elected primary and writes resume. During this time our test infrastructure, simulating an operator, detects the failure and begins preparing a new node, $n_3$, by copying the snapshots onto the new host, $n_3$ tries to send a join message to the primary and succeeds at $B$. At $C$, our test infrastructure starts a governance proposal ($p_3$ in \autoref{lst:reconf}) as consortium member $m_0$ to add $n_3$ to the network and remove $n_0$. Consortium members $m_0$ and $m_1$ then submit ballots to support the proposal $p_3$. By $D$ the governance proposal has been accepted and $n_3$ will now be added to the current configuration at the consensus layer. At $E$, the reconfiguration has been completed, the fault tolerance of the service has been restored, and $n_0$ can be retired by the operator.

\lstset{emph={txid}, emphstyle=\bfseries, emph={[2]map}, emphstyle={[2]\underline}}

\begin{lstfloat}
	\begin{lstlisting}[frame=single,basicstyle=\ttfamily\scriptsize,escapechar=\%]
Bal = "export function vote (proposal, proposer_id) {return true}"
txid 3.198408: %\textcolor{lightgray}{// Point B on \autoref{fig:availability}}%
  map public:ccf.gov.nodes.info:
    %$n_3$%: {status: Pending}
txid 3.207341: %\textcolor{lightgray}{// Point C on \autoref{fig:availability}}%
  map public:ccf.gov.proposals:
    %$p_3$%: {actions: [{name: transition_node_to_trusted, args: {node_id: %$n_3$%}}, {name: remove_node, args: {node_id: %$n_0$%}}]}
  map public:ccf.gov.proposals_info:
    %$p_3$%: {ballots: {}, proposer_id: %$m_0$%, state: Open}
txid 3.208098:
  map public:ccf.gov.proposals_info:
    %$p_3$%: {ballots: {%$m_0$%: Bal}, state: Open}
txid 3.209096: %\textcolor{lightgray}{// Point D on \autoref{fig:availability}}%
  map public:ccf.gov.proposals_info:
    %$p_3$%: {ballots: {%$m_1$%: Bal, %$m_0$%: Bal}, final_votes: {%$m_1$%: true, %$m_0$%: true}, state: Accepted}
  map public:ccf.gov.nodes.info:
    %$n_0$%: {status: Retiring}
    %$n_3$%: {status: Trusted}
txid 3.210844: %\textcolor{lightgray}{// Point E on \autoref{fig:availability}}%
  map public:ccf.gov.nodes.info:
    %$n_0$%: {status: Retired}
\end{lstlisting}
	\caption{Sample of the key updates from the ledger in \autoref{fig:availability}, where $n_0$ is replaced with $n_3$. Adapted for brevity.}\label{lst:reconf}
\end{lstfloat}
\section{Related Work}\label{sec:related}

We divide the state-of-the-art into techniques that do not leverage trusted execution (\autoref{subsec:related/nontee}) and those that do (\autoref{subsec:related/tee}).

\subsection{Approaches without TEEs}\label{subsec:related/nontee}

Raft~\cite{Ongaro14} does not provide confidentiality or integrity on untrusted infrastructure as the protocol only targets high availability. The system model of Raft assumes that nodes can only fail by crashing, and thus does not handle other faults, for instance, nodes subverting the consensus protocol to reorder transactions or fork the ledger. The same is true of other crash fault-tolerant (CFT) consensus algorithms~\cite{Lamport98} and atomic broadcast protocols~\cite{Junqueira11}.

TEEs can help systems to achieve integrity and confidentiality, however, simply executing a CFT protocol within TEEs does not necessarily provide the desired CIA properties. For instance, Engraft~\cite{Wang22} uses model checking to prove that Raft within TEEs does not provide integrity. A key concern is that the persistent storage is outside the trust boundary and thus under the control of the untrusted host, which can truncate the ledger to violate Raft's linearizability guarantees. Moreover, CCF also provides features, such as receipts and multiparty governance, which are beyond the scope of Raft.

Unlike their CFT counterparts, Byzantine fault-tolerant (BFT) protocols, can tolerate arbitrary faults including malicious nodes. However, safety is only guaranteed by BFT protocols provided that a threshold of participants, often two-thirds, are non-faulty. This limitation is fundamental to BFT protocols, yet it is unrealistic for cloud environments where the parties involved are not all running their own nodes. BFT protocols are also typically less available than CFT protocols, as they require at least two-thirds of nodes to be available to make progress and some BFT protocols also do not support reconfiguration, which is key for restoring fault tolerance after a node failure. BFT protocols are computationally expensive relative to CFT protocols and BFT protocols also require the re-execution of user requests across nodes, so the application logic must be deterministic, making it difficult to support general-purpose programming languages. Distributed ledger technologies and permissioned blockchains~\cite{Fabric,Quorum,Corda} often utilize BFT state machine replication to provide distributed multiparty applications with integrity provided that a threshold of participants are non-faulty. Many blockchains struggle with poor performance~\cite{Gramoli23}, high energy footprints~\cite{Platt21}, and typically they do not provide confidentiality, although zero knowledge proofs~\cite{Goldwasser89} may be used to avoid recording private data on blockchains.

Fully homomorphic encryption, arbitrary computation over encrypted data, is an active research area, but it is not yet a practical option~\cite{Martins17} for most applications.
Secure Multiparty Computation provides both confidentiality and integrity but can be expensive and is highly application-specific~\cite{Wang17}.

\subsection{TEE-based approaches}\label{subsec:related/tee}

At the time of writing, there are various TEE platforms, at differing stages of development. These include ARM CCA~\cite{CCA} with TrustZone~\cite{TrustZone} and Realms~\cite{Xupeng22}, Intel SGX~\cite{Costan16} and TDX~\cite{TDX}, RISC-V Keystone~\cite{Lee20}, and AMD SEV~\cite{SEV}, SEV-ES~\cite{SEV-ES}, and SEV-SNP~\cite{SEV-SNP}.
Like most systems discussed in this section, CCF opted for Intel SGX due to its availability at the time of development, SDK support~\cite{OE,IntelSDK,EGo}, and its small TCB relative to VM-based approaches.
Early Intel SGX offerings had highly limited enclave memory (128 MB) but this is no longer the case~\cite{El22}, for instance, the 3rd Gen Intel Xeon Scalable Processors support up to 512 GB~\cite{3rdXeon17}. In January 2022, Intel reaffirmed its commitment to SGX on its server-grade processors~\cite{SGXSupport22}.
There are two general approaches to building trustworthy data systems leveraging TEEs: lift-and-shift and enclave-aware.

Lift-and-shift approaches focus on porting existing applications to a TEE with minimal modifications. Examples include Haven~\cite{Baumann15}, Graphene~\cite{Tsai14,Tsai17}, SCONE~\cite{Arnautov16}, Occlum~\cite{Shen20}, Ryoan~\cite{Hunt16}, Ratel~\cite{Cui22}, S-FaaS~\cite{Alder19}, and Parma~\cite{johnson23}.
Cloud providers are beginning to offer lift-\&-shift as a service, in the form of Confidential VMs or Confidential containers~\cite{Parma}. Kata containers~\cite{Kata} has experimental support for AMD SEV-SNP~\cite{KataCC} and lift-and-shift is already being utilized, for instance, by Constellation~\cite{Constellation}. %
Lift-and-shift is easy to adopt for stateless applications but has a large trusted computing base (TCB), violating the principle of least privilege, and is more challenging for applications requiring durability.

Enclave-aware approaches are new system designs with a focus on dividing code between the trusted enclave and the untrusted host. This approach has a higher barrier to entry compared to lift-\&-shift solutions but also a smaller TCB.
Some systems in this space are highly application specific~\cite{ContactApp, Schuster15, Brenner16, Li21}. Others are more application agnostic, for instance, recent efforts to put privacy into blockchains with off-chain processing on TEEs~\cite{PDO, Xiao20, Cheng19, EthSpec,FPC},
a TEE-based file system~\cite{Kumar21}, or Glamdring~\cite{Lind17} which allows developers to annotate sensitive data and automatically partitions the application between the enclave and host.

\autoref{tab:related} compares CCF to similar TEE-based approaches, along the dimensions of data confidentiality, integrity protection (divided into execution and data), high availability, support for reconfiguration, and support for untrusted infrastructure. All approaches listed are based on Intel SGX, except for Data station~\cite{Xia22} which opted for AMD SEV.

\begin{table}[t]
	\caption{Comparison of data systems utilizing TEEs, ordered by publication year. (C) Confidentiality, (E) Execution Integrity, (D) Data Integrity, (A) High Availability, (R) Reconfiguration, (U) Untrusted infrastructure, (OS) Open Source.}\label{tab:related}
	\begin{tabular}{lllllllll}
		\toprule
		                                      &
		C                                     &
		E                                     &
		D                                     &
		A                                     &
		R                                     &
		U                                     &
		OS                                                                                                                 \\
		\midrule
		Hybster~\cite{Behl17}                 &          & \cmark{} & \cmark{} & \cmark{} &          & \cmark{} &          \\

		Fabric PC~\cite{brandenburger18}      & \cmark{} & \cmark{} & \cmark{} & \cmark{} & \cmark{} &          & \cmark{} \\
		Pesos~\cite{Krahn18}                  & \cmark{} &          & \cmark{} & \cmark{} &          & \cmark{} &          \\
		EnclaveDB~\cite{Priebe18}             & \cmark{} & \cmark{} & \cmark{} &          &          & \cmark{} &          \\
		VeritasDB~\cite{Sinha18}              &          &          & \cmark{} &          &          & \cmark{} &          \\

		Ekiden~\cite{Cheng19}                 & \cmark{} & \cmark{} & \cmark{} & \cmark{} & \cmark{} &          &          \\
		ShieldStore~\cite{Kim19}              & \cmark{} &          & \cmark{} &          &          & \cmark{} & \cmark{} \\
		Speicher~\cite{Bailleu19}             & \cmark{} &          & \cmark{} &          &          & \cmark{} &          \\

		Azure SQL AE~\cite{Antonopoulos2020}  & \cmark{} &          &          & \cmark{} & \cmark{} & \cmark{} &          \\

		Avocado~\cite{Bailleu21}              & \cmark{} &          & \cmark{} & \cmark{} & \cmark{} & \cmark{} & \cmark{} \\
		Enclage~\cite{Sun21}                  & \cmark{} &          &          &          &          & \cmark{} &          \\
		FastVer~\cite{Arasu21}                &          &          & \cmark{} &          &          & \cmark{} &          \\
		HotStuff-M/VABA-M~\cite{Yandamuri21}  &          &          & \cmark{} & \cmark{} &          & \cmark{} &          \\
		Precursor~\cite{Messadi21}			  & \cmark{} &          & \cmark{} &          &          & \cmark{} &          \\

		(Chained) Damysus~\cite{Decouchant22} &          &          & \cmark{} & \cmark{} &          & \cmark{} & \cmark{} \\
		Engraft~\cite{Wang22}                 & \cmark{} & \cmark{} & \cmark{} & \cmark{} &          & \cmark{} & \cmark{} \\
		FlexiTrust~\cite{Gupta22}             &          &          & \cmark{} & \cmark{} &          & \cmark{} &          \\
		Treaty~\cite{Giantsidi22}             & \cmark{} &          & \cmark{} &          &          & \cmark{} & \cmark{} \\
		Data Station~\cite{Xia22}             & \cmark{} & \cmark{} & \cmark{} &          &          & \cmark{} &          \\

		\textbf{CCF}                          & \cmark{} & \cmark{} & \cmark{} & \cmark{} & \cmark{} & \cmark{} & \cmark{} \\
		\bottomrule
	\end{tabular}
\end{table}

Avocado~\cite{Bailleu21} is a confidential, integrity-protected key-value store, built for Intel SGX using SCONE~\cite{Arnautov16}. Unlike CCF, Avocado is purely in-memory with a node replacement protocol based on Hermes~\cite{Katsarakis20} and does not support disaster recovery. Avocado uses multi-writer ABD~\cite{Lynch97} for replication instead of a distributed consensus algorithm. Avocado supports sharding on individual operations since Avocado does not support transactions so no cross-shard transaction support is needed. Like many systems utilizing TEEs, its architecture is heavily influenced by the limited enclave memory in early SGX offerings and the high cost of enclave page cache paging.
Similarly, Treaty~\cite{Giantsidi22} is a confidential, distributed key-value store built on SGX.
The distribution layer is built on top of Speicher~\cite{Bailleu19} and like Avocado, its networking stack is based on eRPC~\cite{Kalia19}.
Treaty uses two-phase commit for replication instead of a consensus algorithm such as Raft. Treaty supports sharding but does not support server-side application logic, so users submit interactive transactions directly. Protection against rollback attacks is achieved using a trusted counter.
FastVer~\cite{Arasu21} adds integrity protection to the Faster~\cite{Chandramouli18} key-value store, using a combination of trusted execution, Merkle trees, and deferred memory verification. Whilst FastVer does not have many of the features of CCF such as confidentiality or custom application logic, it does achieve outstanding performance.
FlexiTrust~\cite{Gupta22} is motivated by the claim that current TEE protocols have three issues compared to classic BFT protocols: liveness attacks, rollback attacks, and sequential consensus. FlexiTrust tries to address these. It is a BFT protocol for TEEs that uses $3f+1$ nodes but only one phase (usually CFT protocols require one phase but BFT protocols require two) and a TEE is only used by the primary. FlexiTrust provides no confidentiality guarantees but does perform well (up to 200K txns/sec).
Hyperledger Fabric~\cite{Fabric} is a permissioned blockchain, whose nodes are divided into peers, which execute the chaincode (smart contract/app logic), and the orderers, which use Raft (implementation from etcd) to order transactions on the ledger. Hyperledger Fabric Private Chaincodes (PC)~\cite{brandenburger18,FPC-RPC} is a proposed extension to Fabric that uses Intel SGX for private execution by modifying the peers to execute chaincode inside an enclave. With Fabric PC, like Ekiden~\cite{Cheng19}, the consensus layer remains outside the TEEs.

We have released a technical report~\cite{Russinovich19} describing an earlier iteration of CCF and a subsequent paper on a technique for providing accountability following TEE compromise~\cite{Shamis22}.
\section{Conclusion \& Future Work}\label{sec:conclusion}

CCF is a general-purpose platform for constructing confidential, integrity protected \& highly available services, with support for delegation to untrusted cloud infrastructure and governance by mutually untrusted parties. CCF is a production quality, open source project that we believe has substantial potential as a foundation for new systems research into trustworthy computing, and it is already being built upon~\cite{PDO,LSKV,did-ccf,scitt}.
CCF offers an all-in-one platform for deploying trustworthy multiparty applications on untrusted infrastructure. Moreover, aspects of CCF's design, such as its disaster recovery procedure, atomic reconfiguration, and governance model, could also be useful techniques in their own right.

CCF is by no means the final word on confidential computing. Our trusted computing base still includes the hardware manufacturer and research is ongoing into verifying the correctness of the code and protecting against side-channel attacks~\cite{Gruss17,Oleksenko18}. Intel SGX has also been the subject of various exploits~\cite{Van20, Van18, SGAxe, Borrello22, Murdock19, Skarlatos19, Schwarz17, VanSchaik22}, capable of compromising the confidentiality and/or integrity of execution. We have applied the available mitigations, however, further exploits may be uncovered.
SGX also has a significant performance overhead, as shown in \autoref{tab:throughput}, and is not universally available. In response, we have recently added experimental support for another trusted execution platform, AMD SEV-SNP, and we aim to support future offerings such as Intel TDX. CCF's threat model is a significant improvement over the current state of the art and a substantial step in the right direction. The removal of cloud providers from the trusted computing base opens up a myriad of new possibilities for reimagining multiparty distributed systems that leverage the cloud's compute capability whilst minimizing trust in cloud providers.

\balance{}

\begin{acks}
	We are sincerely grateful to the following people for their contributions to CCF at various stages of the project: 
	Christine Avanessians,
	Dominic Ayre,
	Jacqueline Aziz,
	Theodore Butler,
	Miguel Castro,
	Mahati Chamarthy,
	Joonwon Choi,
	Renato Golin,
	Istvan Haller,
	Syed Hamza,
	Sid Krishna,
	Sangeeth Kumar,
	Paul Li\'{e}tar,
	Thomas Moscibroda,
	Kartik Nayak,
	Olga Ohrimenko,
	Joe Powell,
	Jonathan Protzenko,
	Maik Riechert,
	Takuro Sato,
	Felix Schuster,
	Roy Schuster,
	Alex Shamis,
	Lyndon Shi,
	Ross P. Smith,
	Cale Teeter,
	Shoko Ueda, and
	Olga Vrousgou.
	We would also like to thank the contributors to Open Enclave~\cite{OE}, Confidential Containers~\cite{johnson23}, and snmalloc~\cite{Lietar19}.
	Finally, we would also like to thank the anonymous VDLB reviewers for their insightful feedback.
\end{acks}

\bibliographystyle{ACM-Reference-Format}
\bibliography{../refs}

\end{document}